\documentclass{aastex}
\usepackage{spr-astr-addons}
\usepackage{url}

\RequirePackage{color}

\newcommand{\emaila}{kuncarayakti@gmail.com}

\begin{document}
\title{On Nova Scorpii 2007 N.1 (V1280 Sco)}

\shorttitle{Nova V1280 Sco 2007}
\shortauthors{Kuncarayakti, Kristyowati, & Kunjaya}

\author{Hanindyo Kuncarayakti} \and \author{Desima Kristyowati}\and \author{Chatief Kunjaya}
\affil{Bosscha Observatory \& Department of Astronomy, Institut Teknologi Bandung, Indonesia}
\email{\emaila}

\begin{abstract}
We present the results of our photometric and spectroscopic observations of Nova Sco 2007 N.1 (V1280 Sco). The photometric data was represented by a single data point in the light curve since the observation was carried out only for one night. The spectra cover two different phases of the object’s evolution during the outburst, i.e. pre-maximum and post-maximum. Measurements of the P-Cygni profile on Na I 'D' line (5889 \AA) was derived as the velocity of shell expansion, yielding $1567.43 \pm 174.14$ km s$^{-1}$. We conclude that V1280 Sco is a fast Fe II-type nova.
\end{abstract}


\section{Introduction}
Nova Sco 2007 N.1 (V1280 Sco) is one of the brightest novae in recent years, since Nova Vel 1999 (c.f. Della Valle et al. 2002). Its discovery was reported by  Yamaoka et al. (2007) and confirmed spectroscopically by Naito \& Narusawa (2007). Found by Nakamura \& Sakurai (c.f. Yamaoka et al. 2007) as a 9$^{\textrm{mag}}$ object on February 4.85 UT, V1280 Sco underwent rapid brightening and eventually reached optical maximum brightness at $\sim 3.7^{\textrm{mag}}$ on February $17^{\textrm{th}}$, and remained so for around three days afterwards. Subsequently it gradually faded and is currently back on its original brightness at $\sim 12^{\textrm{mag}}$. Henden \& Munari (2007) reported its position as $\alpha_{2000.0} = 16^{\textrm{h}}57^{\textrm{m}}41.2^{\textrm{s}}, \delta_{2000.0} = -32^\circ20'35.6"$.

\begin{figure}[h]
    \centering
       \includegraphics[scale=0.55]{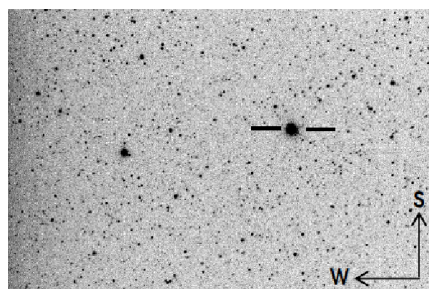}
    \caption{Raw \textit{V}-band CCD image of Nova Sco 2007 N.1. The field was 17' $\times$ 11'. The nova is the brightest star on the image (marked), while the second brightest is SAO 208228. Integration time was 20 seconds.}
\label{fig1}
\end{figure}

\section{Observations}
Our observations were carried out at the Bosscha Observatory, during late February 2007. Due to weather restrictions and observing time availability, we managed to cover only three nights for spectroscopy and one night for photometry. 

On February 13.9 UT we performed a "snapshot" \textit{BVRI} photometry on the object, using a Celestron 11" f/10 Schmidt-Cassegrain mounted with SBIG ST-8XME CCD (1530 $\times$ 1020 px, 9$\mu$ pixelsize). The nearby 7$^{\textrm{mag}}$ star SAO 208228, which was on the same CCD field with V1280 Sco, was used as a comparison star (see Figure \ref{fig1}). 

Spectroscopic observations were carried out on February 16.9, 24.9, and 28.9 UT. We used an SBIG DSS-7 spectrograph combined with the Celestron 11" and ST-8XME (February 16$^{\textrm{th}}$), while on February 24$^{\textrm{th}}$ and 28$^{\textrm{th}}$ the spectrograph was combined with a Celestron 8" f/10 Schmidt-Cassegrain and SBIG ST-7XME-S CCD (765 $\times$ 510 px, 9$\mu$ pixelsize). Due to the absence of internal wavelength comparison source in the spectrograph, we used external sources i.e. skyglow, an Hg lamp, HR 5511, and Arcturus (radial velocity standard star, Udry et al. 1999). We used high resolution spectral atlas of Arcturus from Hinkle et al. (2003).

The combination of our instruments gave a resolution of 5.4 {\AA} per pixel. Effective wavelength coverage was from 3800 to 8000 \AA. Throughout the observations the CCD camera was cooled down to $-$10 C. Integration time was 10 to 20 seconds. Typical sky condition was clear, but not photometric since there were scattered clouds present.

\section{Results}
\subsection{Photometry}
During our 1-hour observing session, we obtained 15 science images for each \textit{BVRI} band, in addition to dark and flatfield frames. Standard reduction of dark-subtracting and flatfielding was employed. We than carried out aperture photometry to both object and comparison on the reduced images using \textsc{Iraf\footnote{\textsc{Iraf} is distributed by the National Optical Astronomy Observatories, which are operated by the Association of Universities for Research in Astronomy, Inc., under cooperative agreement with the National Science Foundation.}/apphot} package.

\begin{figure}[h]
    \centering
       \includegraphics[scale=0.38]{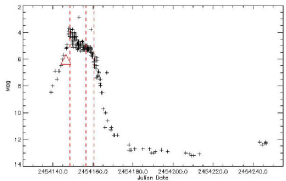}
    \caption{Visual-magnitude light curve of V1280 Sco from validated AAVSO database. Red triangle denotes our \textit{V}-band photometry result on Feb. 13.9, while dashed vertical red lines mark the time of our spectroscopic observations.}
\label{fig2}
\end{figure}

To determine the zero-point of photometry, we refer the \textit{B} and \textit{V} magnitudes of SAO 208228 to the magnitudes from the SAO catalogue. This zero-point was then applied to the instrumental magnitudes of V1280 Sco to transform it into standard magnitudes. Averaging from the 15 frames yielded magnitudes in \textit{B} = 6.400 $\pm$ 0.041 and \textit{V} = 6.011 $\pm$ 0.027. This result is quite in concordance compared to American Amateur Variable Star Observer (AAVSO) database (Figure \ref{fig2}). \textit{R} and \textit{I} magnitudes were calculated by assuming $(V-R)$ = 0.08 and $(R-I)$ = 0.01 \cite{drilling00} for SAO 208228, an A2 star. We obtained for V1280 Sco magnitudes in \textit{R} = 5.702 $\pm$ 0.020 and \textit{I} = 5.420 $\pm$ 0.044.

\subsection{Spectroscopy}
Our spectra were dark-subtracted and flatfield-corrected using \textsc{Iraf}. Spectroscopic reduction was carried out using \textsc{twodspec} and \textsc{onedspec} packages in \textsc{Iraf}. This process resulted in wavelength and flux calibrated spectra.  The fluxes of Feb. 24 spectra were not calibrated since no standard star was observed during the night. The fluxes of Feb 16 and Feb 28 spectra were calibrated using HR 5511 and Arcturus, respectively.  We used calibrated flux values from Hamuy et al. (1992) for HR 5511 and from Breger (1976) for Arcturus.

The spectrum of V1280 Sco on Feb. 16.9, around maximum brightness, showed a stellar-like appearance. The continuum is prominent along with many absorption lines. Balmer lines are apparent but not so strong, and a slight evidence of emission line formation is observable on H$\alpha$. 

\begin{figure}[]
    \centering
       \includegraphics[scale=0.8]{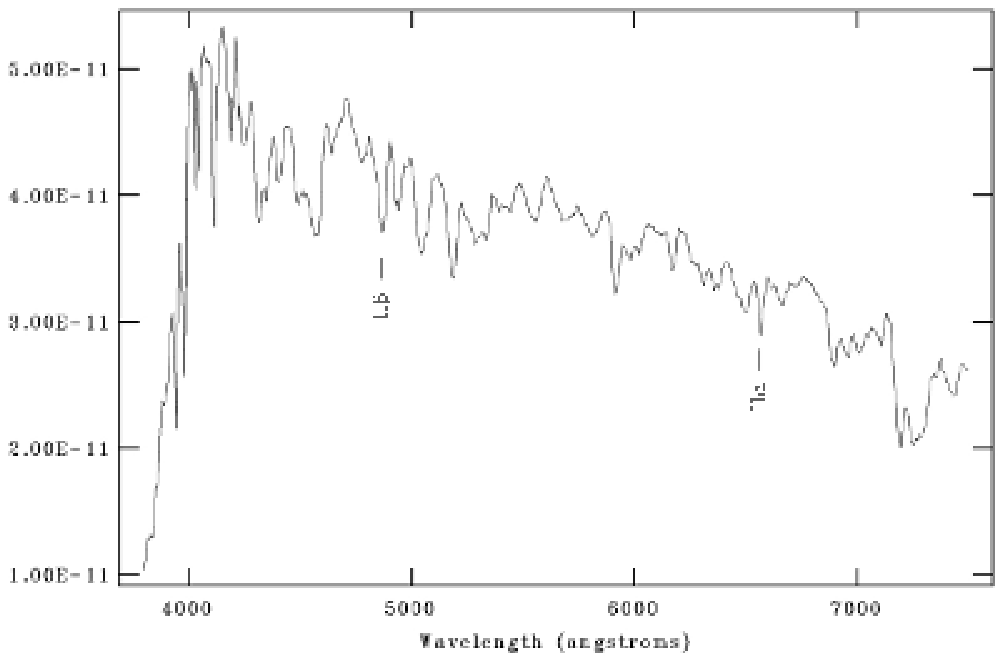}
		\includegraphics[scale=0.75]{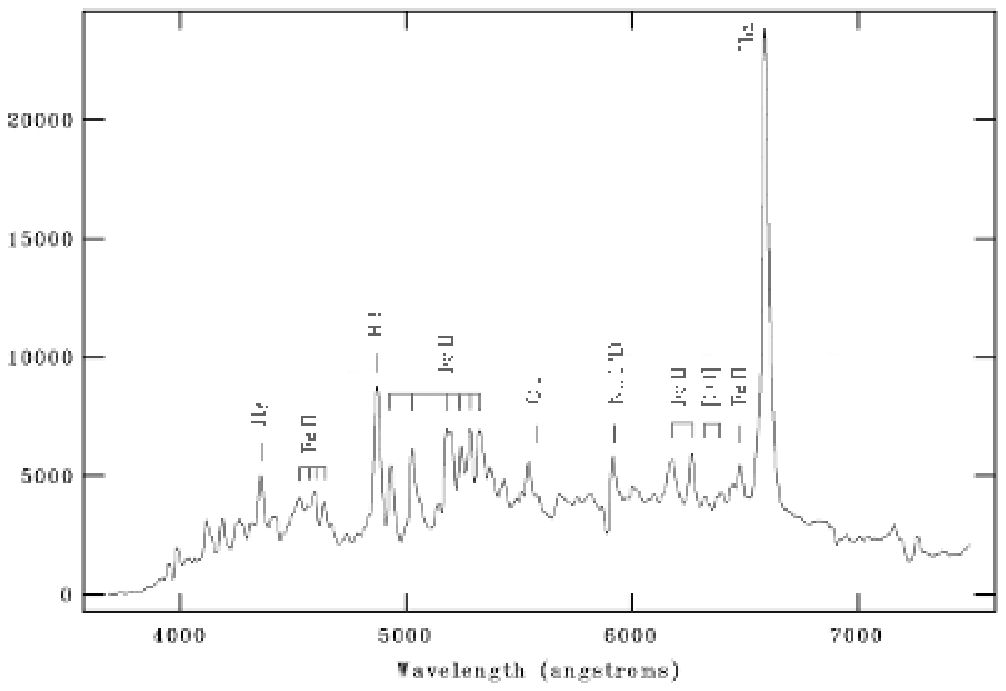}
		\includegraphics[scale=0.8]{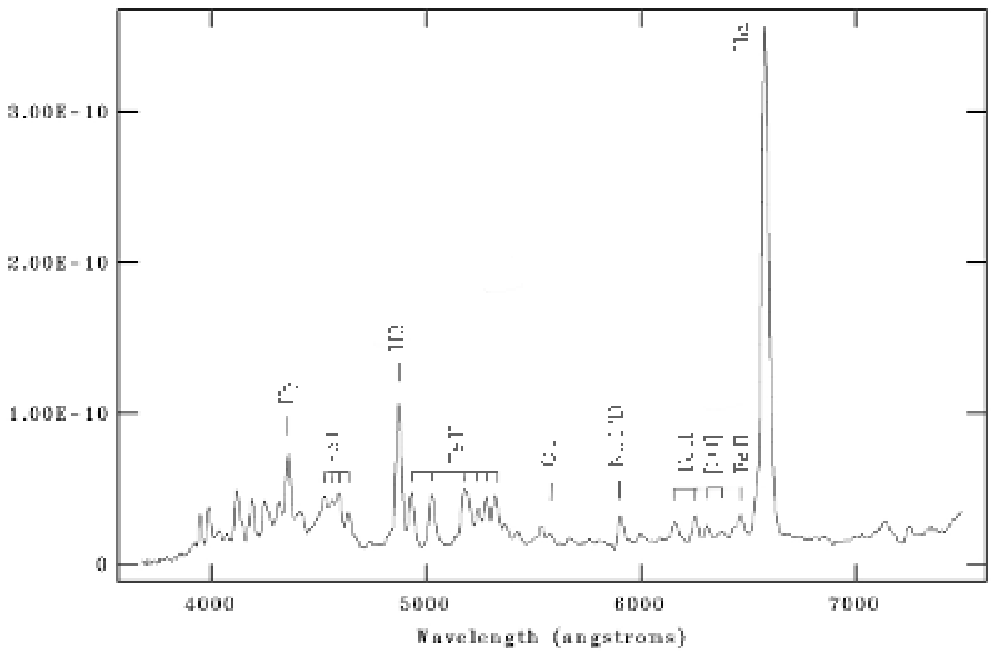}
    \caption{From top to bottom respectively: spectra of V1280 Sco on Feb. 16, Feb. 24, and Feb. 28. Vertical scale is flux in erg cm$^{-2}$ s$^{-1}$ \AA$^{-1}$  (Feb. 16, 28) and counts (Feb. 24).}
\label{fig3}
\end{figure}

Spectra obtained from Feb. 24 and 28, after peak brightness, evidently show a different appearance compared to Feb. 16 spectra. The continuum is less pronounced, while the spectrum is dominated by emission lines. The spectra exhibit strong emissions particularly on H$\alpha$, H$\beta$, and Fe II lines. We detected a P-Cygni profile apparent on around $\lambda$5900 \AA. This feature was identified as Na I (5889 \AA) 'D' line (Figure \ref{fig3}).

To derive shell expansion velocity of the object, we carried out measurements on wavelength of the absorption and emission component of the P-Cyg profile. Prior to this, we normalize the continua of Feb. 24 and Feb. 28 spectra. Measurements to absorption profile were carried out by fitting the profile with a Voigt function, while we used a Gaussian function on the emission component. During the fitting, the base of the fitting function was set at the continuum.

To derive radial velocity $v$ for each Na I 'D' line component from observed wavelength $\lambda_{obs}$ given the rest wavelength $\lambda_0$ and speed of light $c$, we used the Doppler relationship
\begin{equation}
{{\lambda_{obs}-\lambda_0} \over \lambda_0} = {v \over c}.
\end{equation}

Radial velocity derived from this measurement was corrected to radial heliocentric velocity $v_{hel}$. The difference between radial heliocentric velocities of absorption and emission component, i.e. basically difference of wavelength $\Delta\lambda$, yields expansion velocity $v_{exp}$. Measurements of three spectra yielded shell expansion velocity as 1567.43 $\pm$ 174.14 km s$^{\textrm{-1}}$ (Table \ref{table}). 

\begin{table*}
\caption{Measurements on Na I 'D' line. All wavelengths are in {\AA} and velocities are in km s$^{\textrm{-1}}$.}
\begin{tabular}{ccccccc}
\hline Spectrum & $\lambda_{emi}$ & $v_{hel,emi}$ & $\lambda_{abs}$ & $v_{hel,abs}$ & $\Delta\lambda$ & $v_{exp}$ \\
\hline
Feb. 24 \#1 & 5914.03 & 1303.55 & 5884.71 & -189.05 & 29.32 & 1492.59 \\
Feb. 24 \#2 & 5917.76 & 1493.42 & 5889.41 & 50.20 & 28.35 & 1443.22 \\
Feb. 28 & 5902.89 & 796.74 & 5868.19 & -979.54 & 34.70 & 1766.48 \\
\hline
\end{tabular}
\label{table}
\end{table*}

\section{Discussions}
Using the relationship developed by Della Valle \& Livio (1995), we were able to crudely approximate the absolute magnitude of the nova. V1280 Sco took $t_2$ $\approx$ 11 days, i.e. time to decline 2$^{\textrm{mag}}$ from maximum brightness. This decline rate was translated to maximum absolute magnitude $M_V$ = $-$8.63. Interstellar extinction map of Neckel \& Klare (1980) indicates that the amount of extinction in the general direction of V1280 Sco ($l$ = 351.756$^{\circ}$, $b$ = +0.120$^{\circ}$) is $A_V$ $\gtrsim$ 2$^{\textrm{mag}}$. Applying $A_V$ = 2$^{\textrm{mag}}$, and maximum brightness of $V$ = 3.7$^{\textrm{mag}}$, we estimate the distance of the nova as 1.2 kpc. This corresponds to a small vertical distance of the object to the galactic plane, 2.5 pc.

Using [O I] lines at $\lambda\lambda$5577, 6300, 6364 we estimated several physical parameters of V1280 Sco. As described in Ederoclite et al. (2006), the optical depth of $\lambda$6300 line can be calculated from the equation
\begin{equation}
{j_{6300} \over j_{6364}}={{1-e^{-\tau}} \over {1-e^{-\tau/3}}}.
\end{equation}
We found that the value of $\tau_{6300}$ is 5.4, which we used to derive electron temperatures of the ejecta region where [O I] lines originated using equation
\begin{equation}
{T_e} = {{11200} \over {log \left(  {{43\tau} \over {(1-e^{-\tau}})} \times {F_{\lambda6300} \over F_{\lambda5577}}  \right) }}.
\end{equation}
The above equation yield $T_e = 4520$ K. This enabled us to estimate the oxygen mass in the ejecta, in solar mass ($M_\odot$):
\begin{equation}
{M_\textrm{O I}} = 152{d^2_{kpc}}\exp{22850 \over T_e} \times 10^{1.05E(B-V)} {{\tau} \over {1-e^{-\tau}}} F_{\lambda6300}.
\end{equation}
Taking the total-to-selective absorption ratio $A_V/E(B-V) = 3.1$, corresponding to $E(B-V) = 0.65$, we found that the oxygen mass in the ejecta is in the order of $2 \times 10^{-5}M_\odot$. This result, along with the value of $\tau_{6300}$ and $T_e$, is in accordance with the values derived from other novae (c.f. Ederoclite et al. 2006).

According to the rate of brightness drop, V1280 Sco was classified as a fast nova, which has a typical $t_2$ of less than 12 days (Della Valle \& Livio 1998). The derived maximum absolute magnitude and shell expansion velocity were consistent with that of typical fast novae (Gallagher \& Starrfield 1978, Osterbrock 1989).

The spectrum of the nova resembles a "Fe II" type nova spectrum, described in Williams (1992). We classified V1280 Sco then as a fast Fe II-type nova. According to Della Valle \& Livio (1998), the progenitor of this type of nova originated from a relatively old Population I of thin disk/spiral arm and the associated white dwarf is rather massive with mass between 0.9 and 1 $M_{\odot}$.

\acknowledgments
We are very thankful to Dr. H.L. Malasan for both great encouragement and strong criticism. The authors would also like to thank N. Hasanah, P. Irawati, and A.T. Puri Jatmiko for their aid during observations.

\end{document}